\documentclass[%
 reprint,
nofootinbib,
 amsmath,amssymb,
 aps, onecolumn
]{revtex4-2}
\usepackage{graphicx,epsfig}
\usepackage{amsmath}
\usepackage{amsfonts}
\usepackage{braket}
\usepackage{fancyhdr}
\usepackage{color}
\usepackage{comment}

\newcommand{\be}{\begin{equation}}
	\newcommand{\ee}{\end{equation}}
\newcommand{\comp}{{\cal C}}
\newcommand{\g}{g_{r}(\vec{x}_0)}
\usepackage{graphicx}

\begin{document}


\title{The statistics of primordial black holes in a radiation dominated Universe\\ - recent and new results -}

\author{Cristiano Germani}
 \email{germani@icc.ub.edu}
\affiliation{%
 Institut de Ciencies del Cosmos (ICCUB), Universitat de Barcelona, \\Martí i Franquès 1,
E08028 Barcelona, Spain\\
Departement de F\'isica Qu\`antica i Astrofisica, Universitat de Barcelona, Mart\'i i Franqu\`es 1, 08028 Barcelona, Spain
}%
\author{Ravi K. Sheth}
\email{shethrk@physics.upenn.edu}
\affiliation{%
  Department of Physics and Astronomy and Center for Particle Cosmology, \\
  University of Pennsylvania, 209 S. 33rd Street, Philadelphia, PA 19104, USA
}%

\begin{abstract}
We review the non-linear statistics of Primordial Black Holes that form from the collapse of over-densities in a radiation dominated Universe.  We focus on the scenario in which large over-densities are generated by rare and Gaussian curvature perturbations during inflation. As new results, we show that the mass spectrum follows a power law determined by the critical exponent of the self-similar collapse up to a power spectrum dependent cut-off, and that the abundance related to very narrow power spectra is exponentially suppressed. Related to this, we discuss and explicitly show that the Press-Schechter approximation, as well as the statistics of mean profiles, lead to wrong conclusions for the abundance and mass spectrum. Finally, we clarify that the transfer function in the statistics of initial conditions for primordial black holes formation (the abundance) does not play a significant role.
\end{abstract}

\maketitle


\section{The set-up}
\label{sec_intro}

Primordial Black Holes (PBHs), if they exist, are arguably the most economical form of Dark Matter. A PBH is the final state of a large over-density collapse that happened long before matter-radiation equality. A black hole is surely of a primordial origin if its mass is lower than the solar one. In the realm in which PBHs account for all of dark matter, this is precisely where their masses need to be \cite{constraints}.

Assuming only the standard model of particle physics, before matter-radiation equality and after inflation, the Universe was always dominated by radiation. Because black holes only interact gravitationally, they behave like dust particles. In other words, in an homogeneous and isotropic universe (a Friedman-Robertson-Walker Universe, FRW) with metric element
\be
ds^2=-dt^2+a(t)^2 d\vec{x}\cdot d\vec{x}\ ,
\ee
while the energy density of radiation scales like $a^{-4}$, the one of primordial black holes scales like $a^{-3}$. 
In the minimal scenario in which the perturbations to the mean FRW universe were seeded by inflation, large over-densities might only have been generated at scales much smaller than those related to the cosmic microwave background radiation observed today. Thus, for a high reheating temperature, PBHs had a long period to increase their energy density relatively to radiation. In this scenario then, the initial density of PBH was tiny and therefore, the formation of a PBH was a rare event. This particularly resonates with the theory of inflation: inflation mainly generates perturbative over-densities. However, because perturbations are quantum in nature, their amplitudes are statistically distributed and non-perturbative over-densities can be generated too, albeit rarely. 

So far, there is no (perturbative) evidence of deviations from the simple Gaussian statistics, and therefore we shall assume it here: we will discuss the statistics of PBHs formation from a given gaussianly distributed initial seed of curvature perturbations. However, our procedure can be generalized to other distributions.

Before doing so, we need to quickly review the condition for PBH formation in an expanding Universe. While one can certainly consult the review of \cite{albert_rev} for a more technical perspective, here we will give a different, intuitive, view.

In the hoop conjecture of Thorne \cite{hoop}, a black hole forms if a portion of fluid with ``mass'' $M$, in an asymptotically flat spacetime, can be enclosed within a hoop of perimeter $2\pi R_s$ where, $R_s\equiv 2 M$ ($ G_N=\hbar=c=1$) is the Schwarzshild radius of the portion of that fluid. For a fluid of density $\rho$, we can define an instantaneous mass ($dt=0$) as
\be
 M\Big|_{dt=0}\equiv 4\pi\int_0^{R}\rho\ \bar R^2 d\bar R\Big|_{dt=0}\ ,
\ee
where we have assumed that the energy density of the fluid is isotropic and $R$ is the areal radius. In an expanding Universe, this mass is known as the Misner-Sharp mass (MS) \cite{ms}. The scenario we are interested in is a localized large over-density evolving in a homogeneous and isotropic, radiation dominated, Universe \footnote{Strictly speaking the concept of over-density is gauge dependent, we will clarify this later on.}. Because the assumption of asymptotically flatness in the hoop conjecture, we could try to get similar conditions in an expanding Universe by subtracting the infinitely long wavelength mode (the background energy density) from the MS mass. We thus define the instantaneous over-mass 
\be
 \delta M(R)\Big|_{dt=0}\equiv 4\pi \int_0^{R} \delta \rho\ \bar R^2 d\bar R\Big|_{dt=0}\ ,
\ee
where $\delta\rho\equiv\rho-\rho_b$ is the over-density with respect to the background $\rho_b$.
  
The extension of the hoop conjecture to an expanding Universe would then be that a collapse into a black hole starts at position $R_{BH}$ whenever
\be
  {\cal C}\Big|_{dt=0}\equiv \frac{2\delta M(R_{BH})\Big|_{dt=0}}{R_{BH}}\Big|_{dt=0}\sim 1\ .
\ee
We can now remove the fixed time and define
\be
  {\cal C}(r,t)\equiv \frac{8\pi \int_{0}^r\delta \rho\ R(\bar r,t)^2 \partial_{\bar r} R(\bar r,t)\ d\bar r}{R(r,t)}\label{ms}
\ee
where $r$ is the radial coordinate of the general isotropic metric
\be\label{metric}
  ds^2=-A(r,t)^2 dt^2+\frac{\partial_r R(r,t)^2 dr^2}{\Gamma(r,t)^2}+R(r,t)^2d\Omega_2^2\ .
\ee
The metric function $\Gamma$ is the ``boost factor'' of the fluid (being $1$ for vanishing fluid velocity and gravitational mass \cite{ms}) and thus,  the Misner-Sharp mass is the equivalent of the ``rest mass'' of the fluid.  The function $\cal C$, defined in Eq.\eqref{ms}, was called the {\it compaction function} in \cite{shibata}\footnote{In \cite{shibata} the compaction factor was defined without the factor $2$. The factor $2$ was introduced in \cite{ilia} to resemble the Schwarzschild potential. See also \cite{yoo_ss} for the latest interpretations of the earlier work of \cite{shibata}.}. In the same paper, it was shown that, in agreement with hoop conjecture, a black hole would inexorably form whenever ${\cal C}\sim 1$.

Let us now return to our primordial black hole scenario. We have already mentioned that, whenever PBHs are generated by large inflationary perturbations during the radiation epoch, they are generically rare. Assuming a Gaussian distribution of perturbations, this implies an approximate spherical symmetry around the peak of such rare large perturbation \cite{bbks}. Thus, in this regime, the condition for collapse into black holes can be given in terms of $\cal C$:  

The first trapped horizon is formed at a radius equating the maximum of the compaction function \cite{ilia}. In other words, a black hole will form at a scale $r=r_{\rm BH}(t_{\rm BH})$ solution of $\partial_{r}{\cal C}=0$ whenever ${\cal C}(r_{\rm BH},t_{\rm BH})\sim 1$. The threshold for PBH formation will be then given in terms of a critical value $\comp_c(r_m(t_i),t_i)$, where $t_i\neq t_{\rm BH}$ is some initial time \cite{ilia,albert}.

\subsection{Initial conditions and threshold}

The question we shall answer in this section is:
\begin{center}{\it Under what initial conditions for $\cal C$ will a BH form at some later time $t=t_{\rm BH}$?}\end{center}
  Under the inflationary paradigm, perturbations set-in at scales larger than the cosmological horizon.  On those scales, and at leading order in gradient expansion, the metric \eqref{metric} can always be recast into a local FRW metric (see e.g. \cite{diego} for a review)
\be\label{pertmet}
  ds^2_l\simeq-dt_l^2+a_l(t_l)^2 dr_l^2+r_l^2 a_l(t_l)^2 d\Omega^2_2\ ,
\ee
where subscript $l$ refers to local coordinates and functions and $a_l(t)$, the space re-scaling due to a long-wavelength perturbation, i.e. a perturbation with wavelength larger than the cosmological horizon of the background. Up to decaying terms, one finds that $a_l(t_l)\simeq a(t_l)$ where $a$ is the background scale factor in local time \cite{diego}. 
  
  The idea is then to find appropriate initial conditions for $\comp$ at super-horizon scales such that a future black hole would form. At leading order in gradient expansion however, the compaction function vanishes, so it is necessary to go beyond that.  
  
  If the typical comoving scale of the perturbation is $r_0$, we can define the parameter 
\be
  \sigma(t)\equiv\frac{R_H(t)^2}{R_p(t)^2}=\frac{a_i^2}{H^2 a(t)^2r_0^2}= \frac{a_i^2}{\dot a(t)^2 r_0^2}\ll 1\ ,
\ee
where $a_i$ is the scale factor at some initial time $t_i$. Thus $R(t_i,r)=r a(t_i)$.
  
If we specialize to the case of radiation 
\be
 \sigma(t)=\frac{3\, \frac{a(t)^2}{a_i^2}}{(l_p^4\rho_0)\, (r_0/l_p)^2}\ ,
\ee 
where $\rho_0$ is the background energy-density at $t=t_i$ and we have now replaced units by using $l_p$ as the Planck scale. Because the Universe is expanding, a necessary condition is then to choose an initial time such that 
\be
 l_p^4\rho_0\gg \frac{l_p^2}{r_0^2a_i^2}\ .
\ee
As already anticipated, a generic spherically symmetric metric can be recast in the following diagonal form
\be
 ds^2=-A(r,t)^2dt^2+\frac{\partial_r R(r,t)^2}{1-K(r,t)r^2} dr^2+R(r,t)^2d\Omega_2^2\ .\label{metr_exp}
\ee
At super horizon scales one can then expand each function in \eqref{metr_exp} in powers of $\sigma(t)$ \cite{polnarev}. However, it turns out that the zeroth order in $\sigma$ is ambiguous and one might write \cite{polnarev_musco1}
\be
  1-K(r,t) r^2=(1-K_i(r)r^2)\tilde K(r,t)\ ,
\ee
where $K_i(r)$ is the curvature $K$ at an initial time $t=t_i$ where the perturbation is at super-horizon scales, and $\tilde K(r,t)=1+\sum_{n=1}^{\infty}\tilde K_n(r)\sigma(t)^n$. For consistency $K$ should then be next to leading order in gradient expansion and indeed $K_i\sim {\cal O}(l_p^2/r_0^2)$ \cite{polnarev_musco1}, once scales smaller than the horizon are cut-off.
 
With this, at leading order in $\sigma$, but fully non-linearly, one finds that in radiation \cite{ilia, harada0} 
\be
  {\cal C}\simeq \frac{2}{3} K_i(r)r^2\ .
\ee
 As we can see, at super-horizon scales, the compaction function is approximately constant. Thus, a black hole can only be formed once the typical scale of the perturbation re-enters the horizon. This typical scale, as we shall explain, is related to the size $r_m$ in which the compaction function is maximal. Moreover, as it is found numerically \cite{ilia}, the comoving location of this maximum does not change much up to the moment in which the hoop conjecture conditions are met. Thus, we shall approximate it as constant. 
 
 Our prescription for the black hole formation is then that a black hole will eventually form whenever $\comp$ at its maximum is larger than a certain critical value ${\cal C}_c$ that we shall specify later on. The constant(s) $\comp_c$ are called the ``thresholds´´.
 
 The curvature $K_i$ is related to the initial comoving over-density as \cite{ilia}
 \be
 a_i^2 \delta\rho_i(r) \propto (K_i+\frac{r}{3}K_i')\ ,
 \ee
 where, for a function $f$, $f'\equiv\partial_r f$. At the maximum of the compaction function $r=r_m$, where we would like to define a threshold for the black hole formation, one has
 \be
 {\cal C}(r_m)\propto \delta\rho_i(r_m)\,r_m^2\ .
 \ee
 Therefore, the threshold on the compaction function greatly differs from other previous prescriptions which considered a ``threshold'' of the over-density amplitude at its peak $r\sim 0$ \footnote{Here we mean a radius small enough to be closer to the highest point of the over-density but large enough to keep the super-horizon approximation.}. Clearly, $r_m\neq 0$ as $\comp\geq 0$ in the presence of an over-density. Thus, the relation between $\delta\rho(0)$ and $\comp(r_m)$ depends on the full profile of the over-density: any statistics based on the distribution of $\delta\rho_i(0)$ is limited to very specific statistical realizations.  We will return to this point later on.

The initial spatial metric, at time $t=t_i$ and at super-horizon scales, is then
\be
  ds^2\Big|_{t=t_i}\simeq \frac{a_i^2\,dr^2}{1-K_i(r)r^2} + a_i^2\,r^2\,d\Omega_2^2\ 
\ee
(recall that $R(t_i,r) = a_i\,r$).  However, inflationary initial conditions are usually phrased using a different conformal form:  
\be\label{hatr}
  ds^2\Big|_{t=t_i}\simeq a_i^2 \,e^{2\zeta(\hat r)}\left(d\hat r^2+\hat r^2d\Omega^2_2\right)\ ,
\ee
where $\zeta$ is the curvature perturbation in the synchronous gauge. For both expressions to be the same, it must be that
\be
  r = e^{\zeta(\hat r)}\hat r \qquad {\rm and,\ also,}\qquad
 \frac{dr}{\sqrt{1 - K_i(r)r^2}} = e^{\zeta(\hat r)}\,d\hat r.
\ee
However, the expression on the left implies that
 $dr/d\hat r =  e^{\zeta(\hat r)}[1 + \hat r\, \partial_{\hat r}\zeta(\hat r)]$, 
so this, with the expression on the right, implies that 
\be
 K_i(r)r^2 = 1 -\Big[1+\hat r\, \partial_{\hat r}\zeta(\hat r)\Big]^2\ .
\ee  
In the $\hat r$ coordinates, the maximum of the compaction function is at that $\hat r = \hat r_m$ where $\partial_{\hat r }({\hat r}\partial_{\hat r}\zeta)=0$; the amplitude of this maximum cannot exceed $2/3$, due to the regularity condition $K_i(r)r^2<1$.

As for the over-density, it is obvious that the threshold ${\cal C}_c$ cannot generically be written in terms of the value of $\zeta$ at its central peak\footnote{Moreover, because the theory is invariant under shifts of $\zeta$ at super-horizon scales, its peak value only has a meaning by fixing it at some scales. In contrast, the value of the compaction function at its maximum is an observable.}, unless strong assumptions on the statistical realizations of $\zeta$ are made: previous attempts to statistical descriptions of PBH formations, have assumed that the spread of around the mean curvature profile is negligible  \cite{muscoger, yoo}. 
Following \cite{nonlinear} we will, instead, scan over all possible compaction function shapes defining, for each realization, the corresponding threshold.  

\section{Analytic formula for the threshold}

As we have discussed earlier, a gravitational collapse occurs at positions where the maximum of the compaction function is over threshold. The threshold is that value of the compaction function for which a black hole of mass zero is eventually formed. Far away from the maximum, the fluid is dispersed away whereas close to the center, regularity of the curvature $K(r)$, requires the compaction function to decay as $r^2$. Therefore, what matters most for the threshold is the form of the compaction function around its maximum. Moreover, because the conditions to form a black hole are tied to the local gradient pressures, the threshold only approximately depends on the local shape of the compaction function. At the maximum ($r=r_m$), we can then fit the compaction function by using its value ${\cal C}(r_m)$ and its normalized curvature $q\equiv\frac{-r_m^2{\cal C}''(r_m)}{4{\cal C}(r_m)}$ \cite{universal} (see \cite{universal2} for other equations of state).  

Ref.~\cite{universal} used the fitting function
\be
 {\cal C}_{fit}(r)  ={\cal C}(r_m)\ \frac{r^2}{r_m^2}e^{\frac{1}{q}(1-\left[\frac{r}{r_m}\right]^{2q})}\ ,
\ee
where ${\cal C}(r_m)$, $r_m$ and $q$ are calculated from the compaction function of the perturbation.

The average of the fitting function
\be
 \bar {\cal C}(r)=\frac{3}{r_m^3}\theta(r-r_m)\int_0^{r_m} {\cal C}_{fit}(r') r'^2\ dr'
 \label{average}
\ee
would be a fictitious top-hat compaction function which has a threshold, in radiation, equal to $\bar{\cal C}_c=\frac{2}{5}$ \cite{albert}. With this, by inverting \eqref{average}, one obtains the threshold for ${\cal C}(r_m)$ \cite{universal}
\be\label{th}
{\cal C}_c(q)=\frac{4}{15}e^{-\frac{1}{q}}\frac{q^{1-\frac{5}{2q}}}{\Gamma(\frac{5}{2q})-\Gamma(\frac{5}{2q},\frac{1}{q})}\ .
\ee
Note that $\comp_c\geq \frac{2}{5}$; this lower bound has been confirmed by numerical studies.  Numerical simulations show that the critical value \eqref{th} and its dependence on $q$ is accurate to within 2\% \cite{universal}.

If \eqref{th} is given in terms of derivatives with respect to $\hat r$ rather than $r$ \cite[e.g.][]{riotto_r}, then
\be
 q = \frac{-\hat r_m^2\partial_{\hat r}^2{\cal C}(\hat r_m)}{4{\cal C}(\hat r_m)(1-\frac{3}{2}{\cal C}(\hat r_m))}\ .\label{q}
\ee
With this analytical formula for the threshold of PBH formation, we are now in a position to calculate the statistical abundance of PBHs in our Universe.

\section{The statistics of compaction function}

A scalar field (in particular the inflaton) does not have a preferred direction.  Therefore, we expect isolated perturbations to be spherical. This symmetry is broken only by the interference of two or more nearby perturbations \cite{bbks}. Focusing on rare events, as in the case of PBH formed during the radiation era, spherical symmetry around a rare peak of a Gaussianly distributed amplitude, is thus well preserved.

The observed cosmological curvature perturbations at the cosmic microwave background scales (CMB), are extremely close to (multivariate) Gaussian \cite{planck}. We will assume this is true on all scales. Nevertheless, we should warn the reader that, for most of the inflationary evolution related to PBH formation \cite[e.g.][]{prokopec}, this assumption might be too strong \cite[e.g.][]{vicente}.

In the Gaussian case then, as discussed above,
\be
 {\cal C}(\hat r,\vec{x}_0)=\frac{2}{3}\left[1-(1+\hat r\  \partial_{\hat r}\zeta(\hat r,\vec{x}_0))^2\right]\ ,
 \label{eq:C}
 \ee
where $x_0$ defines the {\it center} of the spherical distribution and $\hat r=\lVert\vec{x}-\vec{x_0}\rVert$ in the coordinates \eqref{hatr}. Although the combination $\hat r\  \partial_{\hat r}\zeta(\hat r,\vec{x}_0)$ is independent on the choice of radial units, $\hat r$ is. We will fix the scale factor at matter-radiation equality $a_{eq}=1$ so that, $\hat r$ are physical distances there. 

Before developing the full statistics in details, it is interesting to note that even if $\zeta$ (and thus its derivatives) is multi-Gaussian distributed, $\cal C$ will follow a non-centrally peaked multi-$\chi^2$ distribution. Thus, rare configurations of $\cal C$ generically do not coincide with rare configurations of $\zeta$. 

PBHs are then distributed according to the multi $\chi^2$-statistics of $\cal C$ constrained under the following conditions:
\vspace{0.3cm}

   \paragraph{\bf There is a center} ${\cal C}(\hat r,\vec{x})$ has a peak: there exists a position $\vec{x}_0$ such that 
   	\be
   	\vec{\nabla}{\cal C}(\hat r,\vec{x})\Big|_{\vec{x}=\vec{x}_0}=0\ ,\ {\rm and}\ ,\ \nabla^2{\cal C}(\hat r,\vec{x})\Big|_{\vec{x}=\vec{x}_0}<0\ .
   	\ee 
   	This condition is studied in peaks theory \cite{bbks}. Already at this level, we see that the statistics we are looking for is very different from the Press-Schechter/excursion set approach of Refs.~\cite{ps} and~\cite{ex}, which is often used in the literature.\footnote{Whereas peaks theory seeks to describe the point process which describes the positions around which collapse occurs, the Press-Schechter calculation does {\em not} describe a point process; it only aims at a statistical description of the mass fraction in bound objects.  It returns biased (incorrect) estimates of this mass fraction because it assumes that this can be done by considering the statistics of all positions in space \cite{pramana}, rather than the special subset of positions around which collapse occurs \cite{ps12}.  This point has been discussed extensively in the literature on halo formation during matter domination \cite{pramana,ps12,esp}, and since essentially all of that discussion remains relevant during radiation domination, we will not repeat it here.}  
\vspace{0.3cm}

\paragraph{\bf $\comp(\hat r,\vec{x}_0)$ has a maximum at $\hat r=\hat r_m$} This condition is verified for
\be
\partial_{\hat r}\comp(\hat r,\vec{x}_0)\Big|_{\hat r=\hat r_m}=0\ ,{\rm and}\ ,\ \partial_{\hat r}^2\comp(\hat r,\vec{x}_0)\Big|_{\hat r=\hat r_m}<0\ .
\ee
   This important condition appeared for the first time in the non-linear statistics of \cite{nonlinear} and it is based on the excursion set peaks formalism of Ref.~\cite{esp}. In our statistics we will calculate the probability that, for a given scale $r$ the compaction function has a maximum, thus scanning over all possible realizations of $\comp$. 
   Other proposed statistics \cite{yoo, muscoger} instead, considered only the maximum associated to the averaged profile. This difference is crucial. In our procedure we will be able to associate a threshold for {\it each} statistical realization of the perturbation's shapes (or the above $q$ variable) instead of considering only the statistics of $\comp_c$ for {\it a fixed} (presumably average!) $q$.
   \vspace{0.3cm}
   
   \paragraph{\bf $\comp(\hat r_m,\vec{x}_0)$ is over threshold} The condition is
   \be
   \comp(\hat r_m,\vec{x}_0)\geq \comp_c(q)\ .
   \ee
  Again, this condition crucially differs from those employed in \cite{yoo,muscoger} where thresholds were considered, respectively, for the comoving curvature and the over-density at the peak (i.e. at $r=0$).
  
   \subsection{The statistical variables and the role of transfer function}
   
   As we have already commented, a direct connection to inflation is obtained by considering $\zeta(\vec{x})$ as the main statistical variable. Strictly speaking, this is the curvature perturbation calculated in the past infinity as if radiation was always dominating the Universe evolution \cite{ilia}\footnote{In the language of \cite{ilia}, in the limit of $t\rightarrow 0$.}. Practically, however, numerical simulations start at a finite time from the cosmological singularity. Thus, one would be tempted to consider the statistics of PBHs at the time $t_{BH}$ at which the maximum of the compaction function crosses the cosmological horizon \cite[e.g][]{riotto_r,bellomo} where non-linear effects start to play a crucial role. This statistics would then be developed by hoping that the evolution of $\zeta$, via the perturbative transfer function, would keep invariant its Gaussian nature up to crossing horizon. However, it is expected that the time evolution of curvature perturbations in the non-linear regime, would badly break our assumption of gaussianity. For example, in the context of stochastic inflation, the non-linear regime is no longer Markovian \cite{diego}. 
   
   Even accepting the assumption of gaussianity, because the perturbation at $t=t_{BH}$ is outside the regime of validity of gradient expansion, the use of $\zeta(r_m,t_{BH})$ is {\it not consistent} with the quasi-homogeneous initial conditions of the Misner-Sharp system \cite{polnarev_musco1}, which really use $\zeta(\vec{x})\equiv\zeta(\vec{x}, t\rightarrow 0)$. Thus, using $\zeta(\vec{x},t_{BH})$ would mean to set a different numerical problem from the one studied in the literature \cite{ilia,albert}. In turn, this choice would lead to a background dependent threshold which cannot be obtained as a simple generalization of the one employed here.\footnote{The threshold suggested in \cite{riotto_r} tries to generalize the one of \cite{universal} by using the transfer function in the definition of the compaction function maximum. As we have discussed in the text however, this is inconsistent.} 
    
  Resuming, we will bear with the small error caused by not being able to run the simulation from the infinite past and consider the initial conditions for black hole formation at leading order in gradient expansion. This error is smaller than the error we already accepted by using our analytical formula for the threshold. One can appreciate it by looking at the simulated Hamiltonian constraint deviation caused by the use of quasi-homogeneous conditions at a finite time \cite{albert}. 
  
Having defined the relevant variable, we are now equipped to study its statistical proprieties. In Fourier space
\be\label{fou}
   \zeta(\vec{x})=\int \frac{d^3k}{(2\pi)^3}\  e^{i\vec{k}\cdot\vec{x}}\zeta_k\ ,
\ee
where the $\zeta_k$s modes, assumed to be Gaussian, coincide with the Fourier modes of the curvature perturbations in synchronous gauge and at leading order in gradient expansion. More precisely, the amplitudes follow a Rayleigh distribution, and the phases a uniform distribution.

While the curvature perturbation is a function of $\vec{x}$, $\comp$ is a function of $\hat r$ and $\vec{x}_0$. Thus, we need to specify the meaning of $\hat r\partial_{\hat r}\zeta(\hat r,\vec{x}_0)$: we can integrate the Laplacian of $\zeta$ over a ball $B$ centered on $\vec{x}_0$, obtaining (from now on we will drop the hats)
\be
    \int_B d^3 x\ \nabla^2 \zeta=4\pi r^2\partial_r\zeta(r,\vec{x}_0)\ ,
\ee
where we have assumed spherical symmetry around $\vec{x}_0$ (we have in mind a rare peak from inflation). The same relation can be also written as
\be
     r\partial_r \zeta(r,\vec{x}_0)=\frac{1}{4\pi r}\int d^3x \nabla^2\zeta\  \theta\left(r-\lVert\vec{x}-\vec{x}_0\rVert\right)\ .
\ee 
Using the Fourier decomposition \eqref{fou}, we obtain
\be
     r\partial_r\zeta(r,\vec{x}_0)=-\frac{1}{4\pi r}\int \frac{d^3k}{(2\pi)^3}\ k^2\zeta_k\int d^3x\ e^{i \vec{k}\cdot\vec{x}} \theta\left(r-\lVert\vec{x}-\vec{x}_0\rVert\right)\ .
\ee 
The second integral might be done in polar coordinates
\be
     \int d^3x\ e^{i \vec{k}\cdot\vec{x}} \theta\left(r-\lVert\vec{x}-\vec{x}_0\rVert\right)=2\pi e^{i\vec{k}\cdot \vec{x}_0}\int d\bar r d\phi\ \bar r^2 e^{i k\bar r\cos\phi}\theta\left(r-\lVert\vec{x}-\vec{x}_0\rVert\right)=e^{i\vec{k}\cdot \vec{x}_0}\frac{4\pi}{k^3}\left(-k r\cos(k r)+\sin(kr)\right)\nonumber
\ee
so that we may use the suggestive form
\be
     r\partial_r\zeta=\frac{1}{3} \int \frac{d^3k}{(2\pi)^3}e^{i\vec{k}\cdot \vec{x}_0}  (kr)^2\zeta_k W_{\rm TH} (kr)\ ,
\ee
where we have used the definition of the Fourier transformed top-hat window function
\be
     W_{\rm TH}(kr)=3\,\frac{\sin(kr)-kr\cos(kr)}{(kr)^3}\ .
\ee
     
     We would like to stress here that the top-hat window function {\it was not added by hand}; it is just encoded in the variable $r\partial_r\zeta$.  However, this appearance of $W_{\rm TH}$ ensures that sub-horizon modes of $r\partial_r\zeta$ (for $r$ larger than the cosmological horizon size) are cut-away, a necessary condition for the initial conditions of the Misner-Sharp equations.  Evidently then, to construct the compaction function, we are interested in the Laplacian of the curvature perturbation.

     This Laplacian is also related to the linear density perturbation used in previous statistics \cite{muscoger}. Hence, to gain some intuition, we define the new variable
\be
    g_{r}(\vec{x}_0)\equiv-\frac{4}{3}r\partial_r\zeta=-\frac{4}{9}\int \frac{d^3k}{(2\pi)^3} e^{i\vec{k}\cdot \vec{x}_0} (kr)^2\zeta_k W_{\rm TH} (kr)\ .
\ee
Note that $g_r$ is a linear combination of $\zeta_k$, so it also follows a Gaussian statistics. 

Although the compaction function is {\it fully non-linear}, to gain intuition, we note that $g_r(\vec{x}_0)$ is also the would be smoothed {\it linear} over-density on a ball of radius $r$ at super-horizon scales. This is coming from the Poisson equation of linear perturbations in a FRW Universe. We would like to stress that this linear over-density is just an auxiliary (non-physical) function in the regime we are interested in.

With $g_r(\vec{x}_0)$ we can now write the compaction function as
\be
    \comp(r,\vec{x}_0)=\g\Big(1-\frac{3}{8}\g\Big)\ .
\ee   
Finally, note that, once the condition for $\vec{x}_0$ to be a maximum is imposed, we can shift coordinates and fix, locally, $\vec{x}_0=\vec{0}$.

    \subsection{Statistical conditions I: existence of isolated peaks}

    In what follows, we shall search for a special center $\vec{x}_0=\vec{x}_p$ which is a peak for the compaction function. Indeed, as already discussed, the first constraint to form a PBH is the existence of an isolated peak in the compaction function which will define the spatial center of the approximately spherically symmetric initial over-density (again, we are assuming rare peaks). We need then to restrict the statistical realizations of the compaction function to those that fulfill the constraints
\be
  \vec{\nabla}_{\vec{x}_0}{\comp}(r,\vec{x}_0)=0\qquad {\rm and}\qquad
  \nabla^2_{{\vec{x}_0}}\comp(r,{\vec{x}_0})<0\ .
\ee   
The first condition identifies positions which are extremes, and the second ensures that these extremes are local maxima (i.e., peaks). 

The condition for an extreme of $\comp$, is satisfied either where $\vec{\nabla}_{\vec{x}}g_r({\vec{x}_0}) = 0$ or where $g_r({\vec{x}_0})=4/3$. The latter however has probability zero and would produce a PBH of negligible mass, so we shall discard it. We then see that the condition for the peak of the compaction function is the same condition for an extreme of the linear over-density, as earlier considered in \cite{muscoger}. This extreme is a maximum whenever $0<g_r(\vec{x}_0)<4/3$ and a minimum for $4/3<g_r(\vec{x}_0)<8/3$. The first region corresponds to the so-called type~I collapse and the second to type II \cite{kopp}. Although type~II has yet to be thoroughly explored, it is believed that such peaks collapse rapidly \cite{Harada}. However, because high peaks are extremely rare, and type~IIs are even rarer than type~Is, we will only consider type~Is in what follows (i.e. $g_r(\vec{x}_0)\leq 4/3$).
Moreover, because $g_r$ must exceed $0.49$ for PBH formation (corresponding to the limit $\comp_c\geq 0.4$ \cite{universal}), the conditions for the existence of an (isolated) center of the compaction function are
\begin{equation}
      0.49\lesssim g_r(\vec{x}_p) \leq\frac{4}{3}, \qquad
      \vec{\nabla}_{\vec{x_p}}g_r(\vec{x_p}) = 0\ ,\qquad {\rm and}\qquad
      \nabla^2_{\vec{x}_p}g_r(\vec{x}_p) < 0\ .
\end{equation}
In a discretized sense (see \cite{bbks}), the total number density of peaks is
 \be
 n_{peaks}=\sum_{\vec{x}_p}\delta^{(3)}(\vec{x}-\vec{x}_p)\ ,
 \ee 
the idea is to use the statistical variable $g_r$ and its derivatives instead of the peak(s) position(s).
 
We can expand $g_r$ around the maximum(s) ($\vec{x}_p$)
    \be
    g_r(\vec{x})\simeq g_r(\vec{x}_p)+\frac{1}{2}\nabla_i\nabla_j g_r(\vec{x}_p)(x^i-x^i_p)(x^j-x^j_p)\ .
    \ee
     Defining $\vec{\eta}(\vec{x})\equiv\vec{\nabla}g_r(\vec{x})$ and $\chi_{ij}\equiv\nabla_i\nabla_jg_r(\vec{x_p})$, we have
    \be
    n_{peaks}=\sum_{\vec{x}_p}\Big| {\rm det}\ \chi_{ij}\Big|\delta^{(3)}(\vec{\eta})\ ,
    \ee
    where we have used that around the peak $\eta_i(\vec{x})\simeq \sum_j\chi_{ij}(\vec{x}-\vec{x}_p)^j$.  
    
    The sum over peaks can now be replaced by the probability of having a peak in a position $\vec{x}_p$. As we have already discussed, we assume that $\zeta$, and so $g_r$, is a multi-Gaussian distribution. With this we mean that $\zeta$ is the anti-Fourier transformation of the Gaussian random variables $\zeta_k$. While the probability distribution we look for is a Gaussian on $g_r(\vec{x})$, as we have discussed, concerning the peaks, it is enough to second order expand $g_r$ around the peak value and consider $p(g_r(\vec{x}))\rightarrow p(g_r(\vec{x}_p),\chi_{ij},\vec{\eta}(\vec{x}_p))$.
    
    Statistically, vector, scalar and tensor quantities decouple, thus, $p(g_r,\chi_{ij},\vec{\eta})=p(g_r)p(\chi^T_{ij})p(\chi_r|g_r)p(\vec{\eta})$, where $\chi^T$ is the traceless part of the Hessian matrix of $g_r$. We also defined $\chi_r\equiv -r^2 \nabla^2 g_r(\vec{x}_p)$, which is the trace part of the Hessian. The latter is a scalar and thus correlates with $g_r$. All those probabilities are obviously still multi-Gaussians. 
    
    The mean number of peaks is then
\begin{equation}
    	n_{peaks}\rightarrow\int d\chi_{ij}d\vec{\eta}\ dg_r\Big| {\rm det}\ \chi_{ij}\Big|\delta^{(3)}(\vec{\eta})\theta(\chi_r)\ p(g_r)p(\chi_{ij}^T)p(\chi_r|g_r)p(\vec{\eta})\ ,
\end{equation}
where the Dirac delta and the Heaviside theta define that $\vec{x}_p$ is an extreme that is a maximum.
    
The Dirac delta constraint is easy to integrate:
\begin{equation}
    	\int d\vec{\eta}\ \delta^{(3)}(\vec{\eta})\ p(\vec{\eta})\rightarrow \frac{1}{(2\pi)^{3/2} \sigma_\eta^3} ,
\end{equation}  
where the third power is due to the fact that the distribution is three dimensional and $\sigma_\eta\equiv \sqrt{{\rm det}\left(\langle\eta_i\eta_j\rangle\right)}$.

The integral in $\chi_{ij}^T$ is more involved. However, assuming an approximate rotation invariance (we consider high and rare peaks), the determinant of the Hessian will simply give the normalized trace value to the cube, i.e., $\Big| {\rm det}\ \chi_{ij}\Big|\sim \frac{\chi_r^3}{r^3}$, while the integral over $d\chi^T$ will only give the variance. All in all then 
\begin{equation}
    	\int d\chi_{ij}\Big|{\rm det}\chi_{ij}\Big|\ p(\chi_{ij})\,\theta(\chi_r)\sim \int_0^\infty d\chi_r\,\frac{\sigma_\chi^3\chi_r^3}{r^3}\, p(\chi_r| g_r).
\end{equation}
The exact integration in $\chi^T$ can be actually done exactly leading to 
\begin{equation}
    	n_{peaks}=\int dg_r\int_0^\infty d\chi_r \, \frac{f(\chi_r/\sigma_\chi)}{\left(2\pi r_*^2\right)^{3/2}} \, p(g_r)\, p(\chi_r| g_r)\ ,\label{peaks}
\end{equation}
where the explicit calculation to find $f(x)$ (coming from the integration in $\chi^T$) can be found in appendix A of \cite{bbks}\footnote{The function is
\begin{equation}
	f(x)=\frac{x^3-3x}{2}\left[{\rm erf}\left(\sqrt{\frac{5}{2}}x \right)+{\rm erf}\left(\sqrt{\frac{5}{2}}\frac{x}{2} \right)\right]+\sqrt{\frac{2}{5\pi}}\left[\left(\frac{31x^2}{4}+\frac{8}{5}\right)e^{-5 x^2/8}+\left(\frac{x^2}{2}-\frac{8}{5}\right)e^{-5 x^2/2}\right].
	\end{equation}} and, as anticipated, $f(x)\rightarrow x^3$ for $x\rightarrow \infty$ (large isolated peaks). Finally, $r_*\equiv r\sigma_\eta/\sigma_\chi$.
   
   We pause here for a moment. The integral in $g_r$ has been intentionally left indefinite. The reason is that only a subset of peaks that are maximum of the compaction function and over-threshold will be related to PBHs formation. Because we are looking for rare configurations, we shall assume the existence of only one maximum per given peak, thus discarding the cloud-in-cloud possibility. In this respect, a local maximum would also be a global one. A new condition that $r=r_m$ is a maximum should then be supplemented in the statistics of peaks, which is what we shall do in the next section.

    \subsection{Statistical conditions II: maximum of compaction function}
    
    The second constraint we must impose is the existence of a maximum for the compaction function. Note that here, `maximum' means as $r$ is varied, rather than as the spatial position $\vec{x}$ is varied.  I.e., once the spatial peak position $\vec{x_p}$ has been found, this maximum will now be related to the behavior of the derivatives of $\comp$ with respect to $r$.  Thus, what we really want to count is 
\begin{equation}
    \frac{dn_{peaks\ \cap\  max}}{dr} = \sum_{\vec{x}_p,r_m}\delta(\vec{x}-\vec{x}_p)\delta(r-r_m)\ ,
\end{equation}
where $r_m$ denotes the compaction function maximum for each $\vec{x}_p$, where we have assumed that a peak and a maximum of the compaction function does not happen at more than one smoothing scale.   
    
    As before, for type I black holes, the maxima of $\comp$ coincide with those of $g_r(\vec{x}_0)$. Hence, we can consider the following expansion around $r_m$
\begin{equation}
    \frac{dg_{r}}{dr} = \frac{dg_r}{dr}\Big|_{r_m} + \frac{d^2g_r}{dr^2}\Big|_{r_m}(r-r_m) + \ldots\ = \frac{d^2g_r}{dr^2}\Big|_{r_m}(r-r_m) + \ldots .
\end{equation}
For what follows, it is useful to define the following dimensionless quantities related to the expansion in $r$
\be
	v_r \equiv r\, \frac{dg_r}{dr}, \qquad
	w_r \equiv -r^2\, \frac{d^2g_r}{dr^2},\qquad {\rm and} \qquad
	\chi_r \equiv -r^2 \nabla^2 g_r .
\ee
The curvature of the compaction function is actually related to the Hessian around the peak. Indeed, we have that
\be
  2g_r + w_r = -\frac{4}{9}\int \frac{d^3k}{(2\pi)^3} e^{i\vec{k}\cdot \vec{x}_0} (kr)^4\zeta_k W_{\rm TH} (kr)\ = \chi_r.
\ee
This shows that, for the variables associated with the `infinite' past, $\chi_r$, the (dimensionless) Laplacian of $g_r$, and $w_r$, the second derivative with respect to scale $r$, differ by $2g_r$. I.e., the curvatures with respect to position and scale, $\chi_r$ and $w_r$, differ by $2g_r$.

With the above definitions we have 
\begin{equation}
    v_r = -w_{r_m}\,\frac{(r-r_m)}{r_m}+\ldots .
\end{equation}
and so 
\be
  \delta(r-r_m) = \frac{w}{r_m}\delta(v_r)\sim\frac{w}{r_m}\delta(v_{r_m})
\ee
whenever the Dirac delta is imposed.  The density of those states is then obtained by replacing in \eqref{peaks} 
\begin{equation}
    p(g_r)\,dg_r\rightarrow p(g_{r_m}, v_{r_m}, w)\ dv_{r_m}\, dw\, dg_{r_m}\ . 
\end{equation}
By changing the sum in $r_m$ into an integral, we finally obtain that the number of peaks per unit volume having a maximum at some $r$ is (we now remove for simplicity all the indices $r$)
\begin{equation}
	n_{peaks\ \cap\  max} = \int_{r_{min}}^{r_{max}}\frac{dr}{r}\int_0^\infty d\chi\int_0^\infty dw\ w\int dg\,\frac{f(\chi/\sigma_\chi)}{\left(2\pi r_*^2\right)^{3/2}}\, p(g,v=0,w)\, p(\chi| g,v=0,w)\, \theta(\chi)\ .
\end{equation}
where the integral over only positive values of $w$ specifies that $r_m$ is a maximum.

Because $\chi = 2g + w$, i.e., $\chi$ is completely determined by $g$ and $w$, $p(\chi| g,v=0,w)\rightarrow \delta(\chi-(2g+w))$. Thus, finally we obtain 
\begin{equation}
  n_{peaks\ \cap\  max} = \int_{r_{min}}^{r_{max}}\frac{dr}{r}\int_0^\infty dw\ w\int dg\, \frac{f(\chi/\sigma_\chi)}{\left(2\pi r_*^2\right)^{3/2}}\,p(g,v=0,w)\ ,\label{peakmax}
\end{equation}
where one should read $\chi = 2g + w$. Note that because both $w$ and $g$ are positive, we do not need to add an extra constraint for the maximum in $\vec{x}_p$ (the $\theta(\chi_r)$).

The integral in $r$ requires further explanation. First of all, the minimal radius should be larger than the Hubble radius at initial time for the gradient expansion to be valid. Secondly, because of Hawking evaporation, black holes less massive than $\approx 10^{15}gr$ will be completely evaporated by now and should not be counted \cite{constraints}. Thus, since (replacing Planck units $M_p\simeq 2\times 10^{-5} gr$) $M_\bullet\sim M_p^2 H_r^{-1}=r a(t_r)M_p^2$, we fix 
\be
 r_{min}={\rm MAX}\left[\left(\frac{10^{12}\ gr}{M_p^2}\right)^{3/4} H_{eq}^{-1/4},\left(H(t_i)a(t_i)\right)^{-1}\right]\ ,
\ee
where $H_{eq}$ is the Hubble scale at matter-equality and we used $a_{eq}=1$. Whether one should consider the Hubble radius at initial conditions or the Hawking evaporation limit as a minimal radius, it would depend on the specific inflationary model. 

Similarly, the maximal scale we are interested in is the Horizon size at matter-radiation equality:
\be
r_{max}=H_{eq}^{-1}\ .
\ee
Finally, note that the integral in $r$ is really the integral of all configurations that have a maximum for the compaction function per given smoothing scale $r$. Thus, we will remove from now on the sub-index $m$.

\subsection{Statistics condition III: being over the threshold}

We are now finally in the position of implementing the over-threshold condition for the integral in $g_r$. Conversely to the case of finding a local maximum for the compaction function, the threshold value is in general non-local, i.e. it depends on the profile realization of $g_r$ at all smoothing scales. Thus, the number of peaks we would look for is the subset of $n_{peaks\cap max}$ such that, for any possible smoothing scale configuration {\it with the same peak position and same $r_m$}, the corresponding $g_{r_m}$ is over-threshold. Obviously, this would lead to an untreatable computational problem.

To bypass this issue, previous approaches have considered only the mean $g_r$ profile \cite{muscoger}\footnote{We remind the reader that $g_r$ happens to equal the would be linear over-density \cite{nonlinear}} and thus associated $r_m$ to the location of the compaction function maximum once the mean profile is used. However, unless all possible realizations of $g_r$ for any smoothing scale $r$ have negligible spread from $\langle g_r\rangle$, this peak counting will be grossly wrong (similar argument are for the statistics in $\zeta$ \cite{yoo}). Here we adopt instead a more refined argument already outlined in the introduction:

Although it is true that the threshold at the maximum of the compaction function depends upon the full $g_r$ profile \cite{ilia}, to a very good approximation, the threshold only depends on the curvature of the compaction function around a maximum $r_m$ \cite{universal}. Thus, we could just consider, for any given smoothing scale $r=r_m$, the ensemble of all possible compaction function curvatures and for each one of those associate a threshold.

The realization of this condition requires a bit of algebra: In the region $g\leq \frac{4}{3}$, it is a straightforward exercise to show that \footnote{To show this, note that at the maximum 
	\be
	-r^2\partial^2_r \comp\Big|_{r=r_m}=-r^2\partial^2_r g_r\left(1-\frac{3}{4}g_r\right)\Big|_{r=r_m}=w\sqrt{1-\frac{3}{2}\comp_c}\ .\ee}
\be
 q = \frac{w}{4 \comp_c(q)\sqrt{1-\frac{3}{2}\comp_c(q)}}\ .
\ee
This, together with \eqref{q}, implies $\comp_c(q)=\comp_c(q(w))=\comp_c(w)$.
On the other hand, the threshold in terms of $g$ is
\be
 g_c(w) = \frac{4}{3}\left(1-\sqrt{1-\frac{3}{2}\comp_c(w)}\right)\ .
\ee
Thus, the number density of peaks that would eventually collapse into primordial black holes is
\begin{equation}
	n_{\bullet}=\int_{r_{min}}^{r_{max}}\frac{dr}{r}\int_0^\infty dw\ w\int_{g_c(w)}^{\frac{4}{3}} dg\frac{f(\chi/\sigma_\chi)}{\left(2\pi r_*^2\right)^{3/2}}\,p(g,v=0,w)\ .\label{max}
\end{equation}

\subsection{Energy density in PBHs}

Super-horizon over-threshold perturbations that enter the horizon at the time $r a(t_r)=H(t_r)^{-1}$, quickly collapse into black holes (we shall assume an instantaneous formation) \cite{ilia}. Up to a deviation of about $\Delta\comp\equiv\comp-\comp_c\sim 10^{-2}$ from the threshold value \cite{albert}, the mass is distributed accordingly to the following scaling law \cite{scaling}
\be\label{condm}
 M_\bullet=\frac{{\cal K}}{2 H_r}\,\Bigl[\comp(g) - \comp_c(w)\Bigr]^{0.36}\ ,
\ee 
where $H_r\equiv H(t_r)$. The function $\cal K$ is of ${\cal O}(1)$ and depends on the specific curvature profile chosen, thus, in particular, it does slightly depend on $r$. Nevertheless, we will keep it constant in our estimation for abundances, accepting the ${\cal O}(1)$ error related to it. To fix it, we will take its typical value ${\cal K}\simeq 6$ in \cite{albert}. 

For $\Delta\comp\geq 10^{-2}$, the mass distribution starts to deviate from the scaling law with a maximum of a $\sim 15\%$ deviation \cite{albertenea}. Because the error on using \eqref{condm} is of a similar order as that emerging from fixing $\cal K$, we will stay with it and keep the scaling law all the way through.

Since we are interested in the energy density of PBHs at matter-radiation equality, we need to be careful about the evolution of the number density. So far we have calculated the number density per co-moving volume. Once black holes are formed, as we have already mentioned in the introduction, their energy density simply would dilute as dust, i.e., as the inverse of the volume expansion. At super-horizon scales, and at leading order in gradient expansion, the power spectrum is constant and thus peak positions in co-moving volume will not change, i.e., at leading order, there are no intrinsic velocities between peaks. Obviously, at next to leading order in gradient expansion, this would not be exactly the case. We will not consider further this subtlety here but, we would like to mention that it would actually be incorrect to try to capture the intrinsic velocities between peaks by considering a linear transfer function in the statistics, for similar reasons as those outlined before in the case of threshold definition: The transfer function, capturing the time evolution of the power spectrum at linear order (even assuming that Gaussianity is not broken at next to leading order in gradients), is {\em next} to leading order in gradient expansion. Therefore, considering a transfer function in the statistical correlators would imply the necessity of considering new non-linear thresholds and new non-linear initial conditions. What should be done instead is to calculate the initial number density in co-moving volume as done here and then simulate the velocity dispersion of the initial peaks at later times. We are not aware of any of such numerical simulation so far and thus we shall accept the small error related to consider only the leading order in gradient expansion. 

Having made those remarks, the total energy density of PBHs at matter-radiation equality is
\begin{equation}
 \frac{d^2\rho_\bullet}{dr dM_\bullet}\sim\sum_{\vec{x}_0,r_m,M_\bullet,\comp(r_m)>\comp_c}\delta(\vec{x}-\vec{x}_0)\delta(r-r_m)\ \delta\left(M_\bullet-\frac{{\cal K}}{2 H_r}\,\Bigl[\comp(g) - \comp_c(w)\Bigr]^{0.36}\right)\ M_\bullet 
\end{equation}
or
\begin{equation}
  \rho_\bullet = \int_{r_{min}}^{r_{max}}\frac{dr}{r}\int_0^\infty dw\ w\int_{g_c(w)}^{\frac{4}{3}} dg\frac{f(\chi/\sigma_\chi)}{\left(2\pi r_*^2\right)^{3/2}}p(g,v=0,w)\frac{{\cal K}}{2 H_r}\,\Bigl[\comp(g) - \comp_c(w)\Bigr]^{0.36}\ ,
\end{equation}
where we have again set $a_{eq}=1$.
We can now define the fractional density of PBHs at equality,
 $\beta_\bullet\equiv\frac{\rho_\bullet}{\rho_{eq}}$,
as 
\begin{align}
  \beta_\bullet &= \frac{8\pi}{3H_{eq}^2}\int_{r_{min}}^{r_{max}}
  \frac{dr}{r}\int_0^\infty dw\, w \int_{g_c(w)}^{\frac{4}{3}} dg\,
  \frac{f(\chi/\sigma_\chi)}{\left(2\pi r_*^2\right)^{3/2}}\,
  p(g,v=0,w)\,\frac{{\cal K}}{2 H_r}\,\Bigl[\comp(g) - \comp_c(w)\Bigr]^{0.36}\nonumber\\
  &= \frac{4\pi\,{\cal K}}{3(H_{eq}r_{eq})^3} 
  \int_{r_{min}}^{r_{max}} \frac{dr}{r}\,\frac{(r/r_{eq})^2}{(r/r_{eq})^3}\,
  \int_0^\infty dw\, w \int_{g_c(w)}^{\frac{4}{3}} dg\,
  \frac{f(\chi/\sigma_\chi)}{\left[2\pi (r_*/r)^2\right]^{3/2}}\,
  p(g,v=0,w)\,\Bigl[\comp(g) - \comp_c(w)\Bigr]^{0.36}\ .
  \label{eq:betaBH}
\end{align}
(Recall that $H_{eq}r_{eq} = a_{eq}^{-1}$, and $a_{eq}=1$ in our units.) 

The final expression makes it easy to see that the abundance of PBHs with respect to radiation grows with the scale factor from formation.  If we forget the scaling solution and only focus on order of magnitudes, i.e we consider that the PBH mass is simply given by the mass within the horizon at formation, then we can ignore the integrals over $w$ and $g$ in the mass while still keeping the $r$ dependence. Then, because ${\cal K}/(2H_r)\propto r^2$ and $(2\pi r_*^2)^{3/2}\propto r^3$ we find that, for each $r$ we get $r_{\rm eq}/r=a_{eq}/a(t_r)$.  This is the scaling we should expect:  the integrals over $w$ and $g$ serve to return the actual fraction of patches which produce PBHs.  And the fact that they depend on $r$ shows how the energy density that is stored in PBHs builds up over time (i.e. as $r$ increases). 

\subsection{Mass spectrum of PBHs}

The mass distribution per proper volume (also known as the `mass function') is 
\begin{equation}
 \frac{dn_\bullet(M_\bullet)}{dM_\bullet} \sim \sum_{\vec{x}_0,r_m,\comp(r_m)>\comp_c}\delta(\vec{x}-\vec{x}_0)\delta(r-r_m)\ \delta\left(M_\bullet-\frac{{\cal K}}{2 H_r}\,\Bigl[\comp(g) - \comp_c(w)\Bigr]^{0.36}\right)
\end{equation}
leading to
\begin{equation}
 \frac{dn(M_\bullet)}{dM_\bullet} = \int_{r_{\rm min}}^{r_{\rm max}} \frac{dr}{r} 
	\int_0^\infty \!\!\! dw\, w \int_{g_{c}(w)}^{4/3} dg\,\frac{p(g,w,v=0)\,f(x)}{(2\pi r_*^2)^{3/2}} \
       \delta_{\rm D}\left(M_\bullet - {\cal K} M_{\rm H} [C(g) - C_c(w)]^{0.36}\right)
        \label{eq:nmBH}
\end{equation}
where $g_{c}(w) = (4/3) \, [1 - \sqrt{1 - 3C_c(w)/2}]$ and we use $x\equiv \frac{\chi}{\sigma_\chi}$. 

We can also write the previous expression in a different form:
\begin{equation}
 \frac{dn(M_\bullet)}{dM_\bullet} =  \int \frac{dr}{r} \,
 \int_0^\infty \!\!\! \frac{dw}{w}\,\frac{d^3n(M_\bullet,r,w,v=0)}{d\ln r\ d\ln w\ dM_\bullet} ,\label{total}
\end{equation}
where
\begin{equation}
  \frac{d^3n(M_\bullet,r,w,v=0)}{d\ln r\ d\ln w\ dM_\bullet}
  \equiv \int_{g_{c}(w)}^{4/3} dg\,\frac{w\,p(g,w,v=0)\,f(x)}{(2\pi r_*^2)^{3/2}}\
  \delta_{\rm D}\left(M_\bullet - {\cal K} M_{\rm H} [C(g) - C_c(w)]^{0.36}\right).\label{n3}
\end{equation}
The delta-function suggests that we define
\begin{equation}
  M_g \equiv {\cal K} M_{\rm H} [C(g) - C_c(w)]^{0.36}
  = {\cal K} M_{\rm H} (2/3)^{0.36} [1 - \tilde{C_c}(w) - (1 - \tilde{g})^2]^{0.36} ,
\end{equation}
where we have also defined $\tilde{g}\equiv g/(4/3)$ and $\tilde{C}\equiv C/(2/3)$ (i.e., both quantities in units of their maximum possible value) and $M_H\equiv 1/(2 H_r)$.  Note that $M_g = 0$ when $g=g_{c}(w)$ and $M_g = [1 - \tilde{C_c}(w)]^{0.36}$ when $g=4/3$.  Inverting to write $g$ as a function of $M_g$ yields
\begin{equation}
  g = \frac{4}{3} \left[1 - \sqrt{1 - \tilde{C_c}(w) - \frac{3}{2}\left(\frac{M_g}{{\cal K}M_{\rm H}}\right)^{1/0.36}}\right] ,
 \label{eq:gbullet}
\end{equation}
so
\begin{equation}
  \frac{dg}{dM_g}\Big|_{w\,{\rm fixed}} = \frac{2}{3} \frac{(3/2) (M_g/{\cal K}M_{\rm H})^{1/0.36}/0.36M_g}{\sqrt{1 - \tilde{C_c}(w) - \frac{3}{2}\left(\frac{M_g}{{\cal K}M_{\rm H}}\right)^{1/0.36}}}
  = \frac{(M_g/{\cal K}M_{\rm H})^{1/0.36}/0.36M_g}{1 - \tilde{g}} .
\end{equation}
As a result, instead of the original integral in $dg$ at fixed $w$ in \eqref{n3}, we can consider an integral in $dM_g$. By defining $g_\bullet$ such that $M_g=M_\bullet$, we get
\begin{equation}
  \frac{d^3n(M_\bullet,r,w,v=0)}{d\ln r\ d\ln w\ d\ln M_\bullet} =
	\frac{(M_\bullet/{\cal K}M_{\rm H})^{1/0.36}/0.36}{\sqrt{1 - \frac{3C_c(w)}{2} - \frac{3}{2}\left(\frac{M_\bullet}{{\cal K}M_{\rm H}}\right)^{1/0.36}}}\,
	\frac{w\,p(g_\bullet ,w,v=0)\,f(x_\bullet)}{(2\pi r_*^2)^{3/2}}
 \label{eq:nmScaling}
\end{equation}
(recall that, for the type I black holes, the term in the square-root is always positive).

The $r$ dependence in this expression arises from the $r$-dependence of the various correlators (which we describe shortly), and also because $M_{\rm H} = M_{\rm eq}\,(r/r_{\rm eq})^2$.  This shows that the power law $(M_\bullet/{\cal K}M_{\rm H})^{1/0.36}$, which derives from the scaling solution, is a generic prediction of our approach\footnote{See a similar result for the case of bubbles formation \cite{jaume} where, similarly, the power law in the mass spectrum comes from a Jacobian.}.  The question is:  Do the other terms, and the subsequent integrations over $w$ and $r$, modify this power law?
Before we answer it, we would like to stress once more that our approach, equation~(\ref{eq:nmBH}), explicitly integrates over PBH profile shapes (parametrized by $w$) and PBH formation times (parametrized by $r$); it does {\em not} assume that either of these distributions are sharply peaked.  

For notational convenience, it is convenient to define
\be
   {N}_{rw}(M_\bullet)\equiv \frac{d^3n(M_\bullet,r,w,v=0)}{d\ln r\ d\ln w\ d\ln M_\bullet},
   \qquad
 {N}_{r}(M_\bullet)\equiv \frac{d^2n(M_\bullet,r,v=0)}{d\ln r\ d\ln M_\bullet}
 \qquad {\rm and}\qquad 
 {N}(M_\bullet)\equiv \frac{dn(M_\bullet,v=0)}{d\ln M_\bullet},
\ee 
where the second expression is the result of integrating the first over $w$, the final is the result of integrating over $r$ as well, and it is understood that we always have $v=0$.  Thus,
\be
    {N}(M_\bullet)
    = \int_{r_{\rm min}}^{r_{\rm max}}\frac{dr}{r}\,{ N}_{r}(M_\bullet)
    = \int_{r_{\rm min}}^{r_{\rm max}}\frac{dr}{r}\int_0^\infty \frac{dw}{w}\,{ N}_{rw}(M_\bullet).
 \label{eq:NrNrm}
\ee
If we think of ${N}(M_\bullet)$ as the final PBH `mass function' (i.e. at equality), then it is the sum of all PBHs formed at earlier times, indexed by $r$, ${ N}_{r}(M_\bullet)$; at any fixed time there can have been a range of compaction function `shapes' indexed by $w$, and ${N}_{rw}(M_\bullet)$ is the mass function at fixed $r$ and $w$.

\subsection{The probability distribution}
    Now that we have implemented the constraints, we need to specify the probability distribution $p(g,v=0,w)$.  It will be useful to use the identities
\be
 p(g,v=0,w) = p(v=0)\,p(g|v=0)\,p(w|g, v=0) = p(v=0)\,p(w|v=0)\,p(g|w, v=0),
 \label{eq:bayes}
\ee
where, e.g., $p(w|g, v=0)$ is the conditional probability of having $w$ given $g$ and $v=0$. Because the $\zeta_k$ are Gaussian by assumption, all the other variables are too and thus we only need to consider two-point correlators. Defining
\be
  \sigma_j^2(r) = \frac{16}{81}
    \int \frac{dk}{k}\,(kr)^4\, {\cal P}_\zeta(k)\, W_{\rm TH}^2(kr)\,(kr)^{2j}
  \qquad {\rm where}\qquad
  {\cal P}_\zeta(k) = \frac{k^3\,\langle\zeta_k\zeta_{k'}\rangle}{2\pi^2}\,
    \delta^{(3)}(k+k') ,
\ee
one has 
\begin{align}
    	\sigma_g^2 &\equiv \langle g^2\rangle = \sigma_0^2,\qquad \langle \eta_i\eta_j\rangle=\frac{\sigma_1^2}{3}\delta_{ij},\qquad
    	\sigma_\chi^2 \equiv \langle \chi^2\rangle = \sigma_2^2\ ,\qquad
    	\sigma_w^2 \equiv \langle w^2\rangle = \sigma_2^2 - 4\sigma_1^2 + 4\sigma_0^2,
    	\nonumber\\
    	\sigma_v^2 &\equiv \langle v^2\rangle
    	= \frac{d\langle gv\rangle}{d\ln r} - \langle gv\rangle + \sigma_1^2 - 2\sigma_0^2, \qquad {\rm where}\qquad
    	\langle gv\rangle=\frac{1}{2}\frac{d\sigma_0^2}{d\ln r},
\end{align}
and
\begin{align}
    \label{gammas}
    \langle g\chi\rangle &=\sigma_1^2\ ,\qquad
    \langle gw\rangle=\sigma_1^2-2\sigma_0^2\ ,\qquad
    \langle w\chi\rangle=\sigma_2^2-2\sigma_1^2\ ,\qquad
    \nonumber\\
    \langle v \chi\rangle &= \frac{1}{2}\frac{d\sigma_1^2}{d\ln r}-\sigma_1^2\ ,
    \qquad
    \langle v w\rangle=\langle v \chi\rangle-2\langle v g\rangle.
\end{align}
With these correlators in hand, we can now write:    
\be
    p(w,v=0) = \frac{e^{-\frac{w^2}{2\sigma_w^2(1-\gamma_{wv}^2)}}}{\sqrt{4\pi^2\sigma_v^2\sigma_w^2(1-\gamma_{wv}^2)}}\qquad {\rm and}\qquad
    p(g|w,v=0) = \frac{e^\frac{-(g-\langle g|v=0,w\rangle)^2}{2\Sigma^2_{g|vw}}}{\sqrt{2\pi\Sigma^2_{g|vw}}}\ ,
\ee
where we used the fact that $p(v=0) = 1/\sqrt{2\pi\sigma_v^2}$, and we have 
\be
 \langle g|v=0,w\rangle = \sigma_g\,
 \frac{\gamma_{wg} - \gamma_{wv}\gamma_{vg}}{1 - \gamma_{wv}^2}\,\frac{w}{\sigma_w}
\ee
and
\be
 \Sigma^2_{g|wv} = \sigma_g^2\,
 \frac{1 - \gamma_{gv}^2 - \gamma_{wv}^2 -\gamma_{wg}^2 + 2\gamma_{gv}\gamma_{wv} \gamma_{wg}}{1 - \gamma_{wv}^2}.
\ee
Note that the probability $p(g|w,v=0)$ is {\it not} centered on $g=0$ due to its conditional nature and that we have used the normalized (Pearson) correlation coefficient $\gamma_{ab}\equiv \langle ab\rangle/\sigma_a\sigma_b$.  These, unless ${\cal P}_{\zeta}$ is a power law, also depend on $r$. 

Before we consider explicit examples, as $p(g|w,v=0)$ is a Gaussian, we expect the mass function ${ N}_{rw}(M_\bullet)$ to be a power-law in $M_\bullet$ times a Gaussian cutoff.  Note that this cutoff is not simply $\exp[-(M_\bullet/M_f)^{1/0.36}]$ as suggested in \cite{yokoyama1998} by the use of Press-Schechter formalism, both because $g_\bullet$ is a more complicated function of $M_\bullet$ (see equation~\ref{eq:gbullet}), and because $p(g_\bullet|w,v=0)$ is not centered on zero. Once more this shows that Press-Schechter formalism is not adequate for PBHs. 

\section{Illustrative examples}
We now consider some examples of what our approach implies. It will be useful for the following to note that $\sigma_j\propto r^j$ for sufficiently large $r$; this is a consequence of the built-in top hat filter, and the fact that we do not employ an additional transfer function when computing statistics \cite[for more discussion, see Ref.][]{nonlinear}.

\subsubsection{Sharp feature}
Previous work assumes that if 
\be
   {\cal P}_{\rm sharp}(k)=A_s\, k_p\, \delta(k-k_p),
\ee
then this will produce a PBHs with a well-defined mass.  In this case, the idea is to set $k_p$ so as to produce approximately asteroid mass objects, and $A_s$ is set by requiring that their abundance accounts for all the dark matter.

However, upon setting $\kappa\equiv k_p r$, we have 
\be
 \sigma_j^2(r) = A_s\,\kappa^4\,W^2(\kappa)\,\kappa^{2j}\qquad
 \sigma_w^2(r) = (\kappa^2 - 2)^2\,\sigma_0^2 \qquad
 \sigma_v^2(r) = A_s\,\kappa^4\,\Big(3j_0(\kappa) - W(\kappa)\Big)^2
\ee
and
\be
 \langle gw\rangle = \sigma_0\sigma_w\qquad
 \langle wv\rangle = \sigma_w\sigma_v\qquad
 \langle gv\rangle = \sigma_0\sigma_v .
\ee
As a result, $\gamma_{gv}=\gamma_{wv}=\gamma_{gw}=1$, suggesting that $g$, $v$ and $w$ only differ from one another by multiplicative factors.  Hence, as we are interested in $v=0$, the others are also peaked to zero.  But $g=0$ does not make any PBHs at all! 

While this power spectrum relates the linear with the non-linear statistics (see \cite{nonlinear}), in the non-linear case the contribution of PBHs from a very peaked power spectrum is exponentially suppressed. This does not happen for statistics using the mean profiles \cite{yoo,muscoger} because there, the constraint of having a maximum in the compaction function ($v=0$) does not enter in the statistics.

\subsubsection{Broad feature}
Finally, we consider a model that has power over a broad range of scales, before cutting exponentially at $kr_p\gg 1$: 
\be
   {\cal P}_{\rm broad}(k)=A_s\, (kr_p)^p\, \exp(-kr_p).
\ee
On physical grounds, we expect $p\le 4$ \cite{tasinato}.  We will discuss the values of $A_s$ and $k_p$ shortly.  

The appearance of $r_p$ means that it is more convenient to express our results in terms of dimensionless quantities.  E.g., because of the factor of $r_{eq}/r$ inside the integral in the second of equations~(\ref{eq:betaBH}), it is better to work with
\be
 \tilde{\beta} \equiv (r_p/{\cal K}r_{eq})\, \beta_\bullet 
 \label{tildeBeta}
\ee
rather than $\beta_\bullet$, the mass fraction at equality, itself.  
Likewise, it is better to work with scaled number densities 
\be
  \tilde{N}(M_\bullet) \equiv \frac{4\pi\,r_p^3}{3}\, { N}(M_\bullet)
 = \frac{{ N}(M_\bullet)}{\rho_{eq}/{\cal K}M_p}\, \frac{r_p}{{\cal K}r_{eq}};
 \label{tildeN}
\ee
the first equality is dimensionless (a volume times a number density) and the final expression shows that $\tilde{N}$ scales the number density by $(\rho_{eq}/{\cal K}m_p)$, so the remaining factors are the same as those which arise when defining $\tilde{\beta_\bullet}$.

\vspace{0.2cm}

\begin{figure}
 \includegraphics[width=0.475\linewidth]{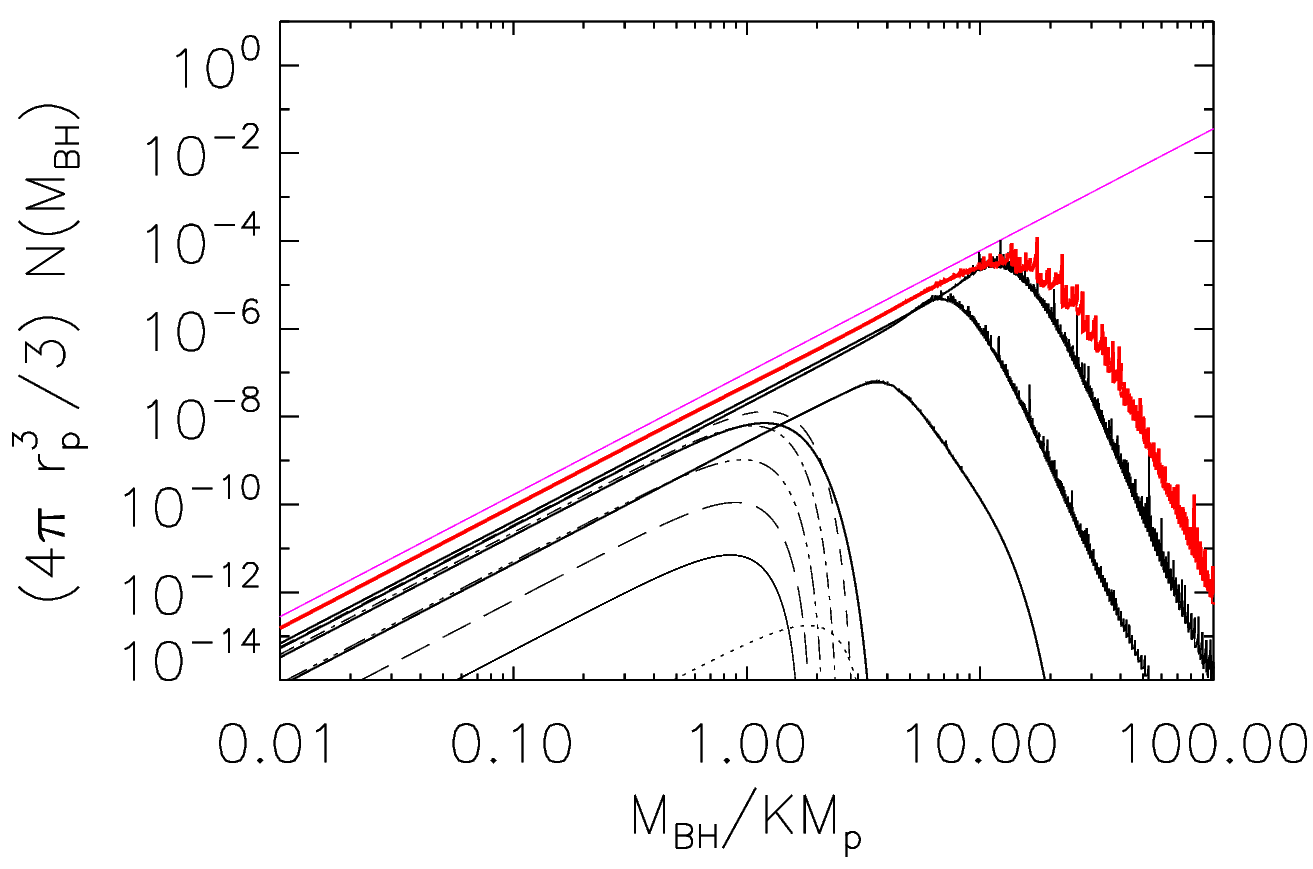} 
 \includegraphics[width=0.475\linewidth]{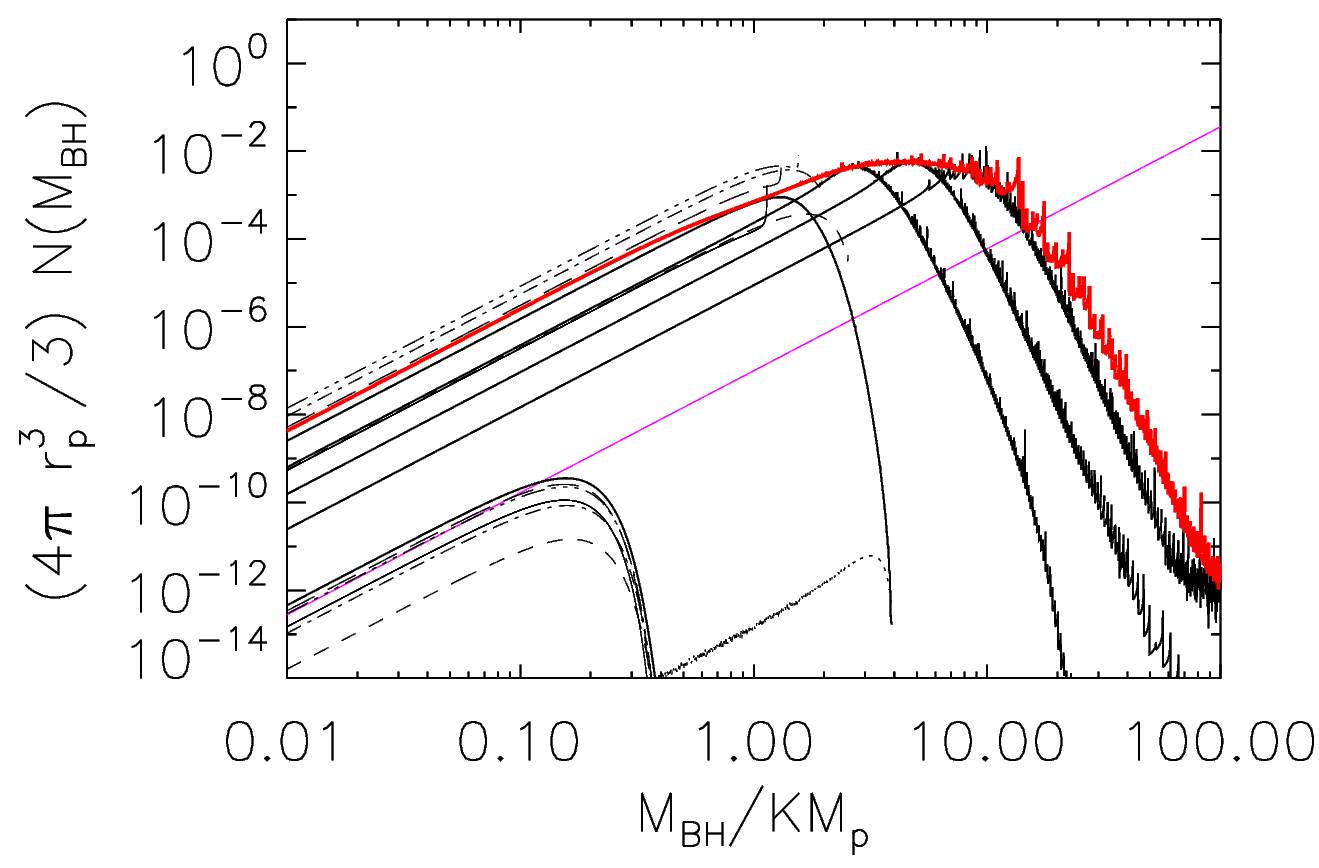} 
 \caption{Distribution of $M_\bullet$ when ${\cal P}(k)$ has slope $p=0$ and amplitude $A_s$ such that $\sigma_0 = 0.33$ (left) and 0.6 (right).  Thick red curve in each panel shows the scaled mass function $\tilde{N}(M_\bullet)$ (equation~\ref{tildeN}); it results from summing over $\tilde{N}_r(M_\bullet)$ distributions, some of which are shown as thick black curves (larger $r$ extend to larger $M_\bullet$).  Each of these results from summing over different $\tilde{N}_{rw}(M_\bullet)$ (cf. equation~\ref{eq:NrNrm}), which we show for a few representative values of $w$ (dotted, short-dashed, dot-dashed, dot-dot-dot-dashed, long dashed).  At small $r$, increasing $w$ decreases the maximum mass.  At larger $r$, the mass function is a power-law with a small divergence at the largest allowed masses:  the amplitude of this power-law depends on $w$, and qualitatively traces the distribution of $w$ (i.e., it is small at both small and large $w$).  Magenta line shows a power-law of slope $1/0.36$ which the text argues show be a good approximation at small $M_\bullet$.\\}
 \label{fig:fixedR0}
 \includegraphics[width=0.475\linewidth]{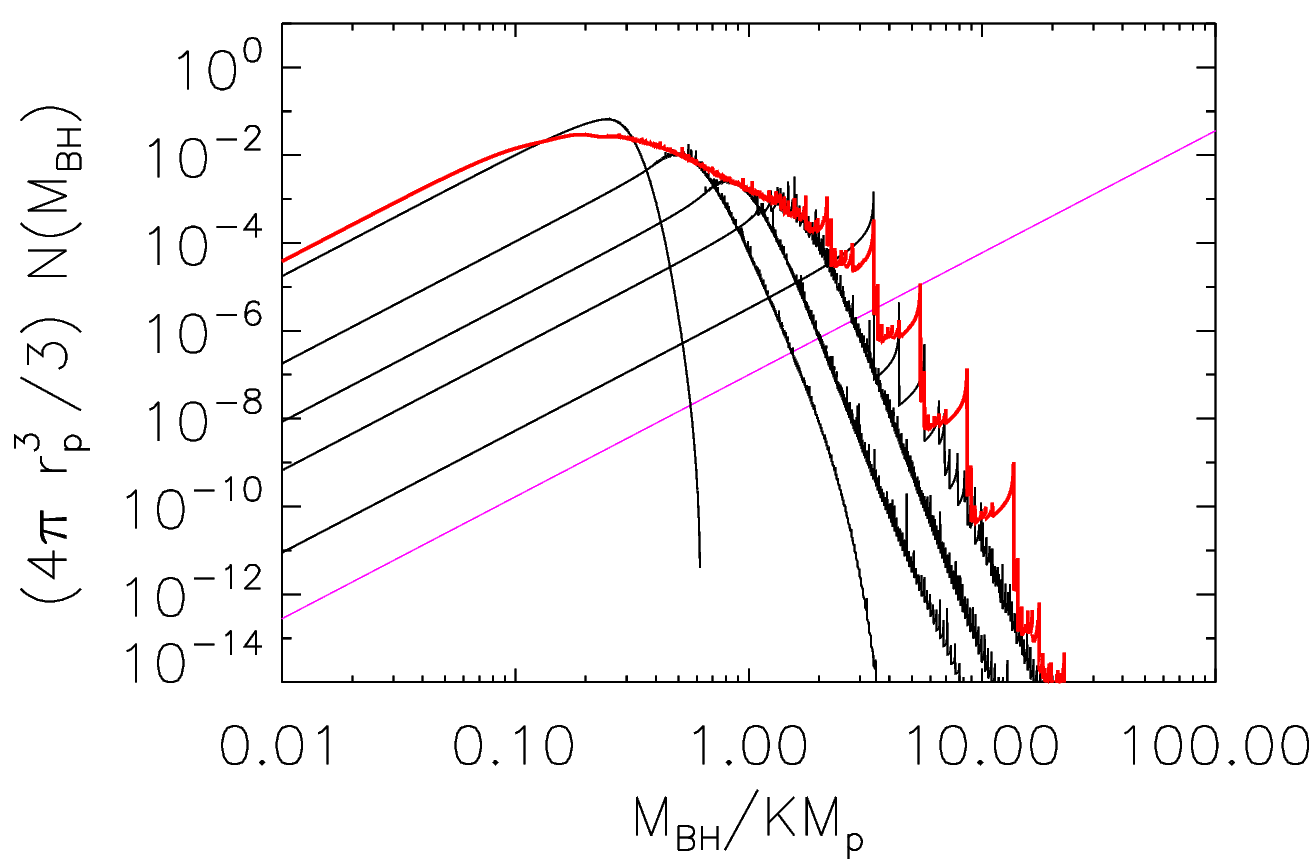} 
 \includegraphics[width=0.475\linewidth]{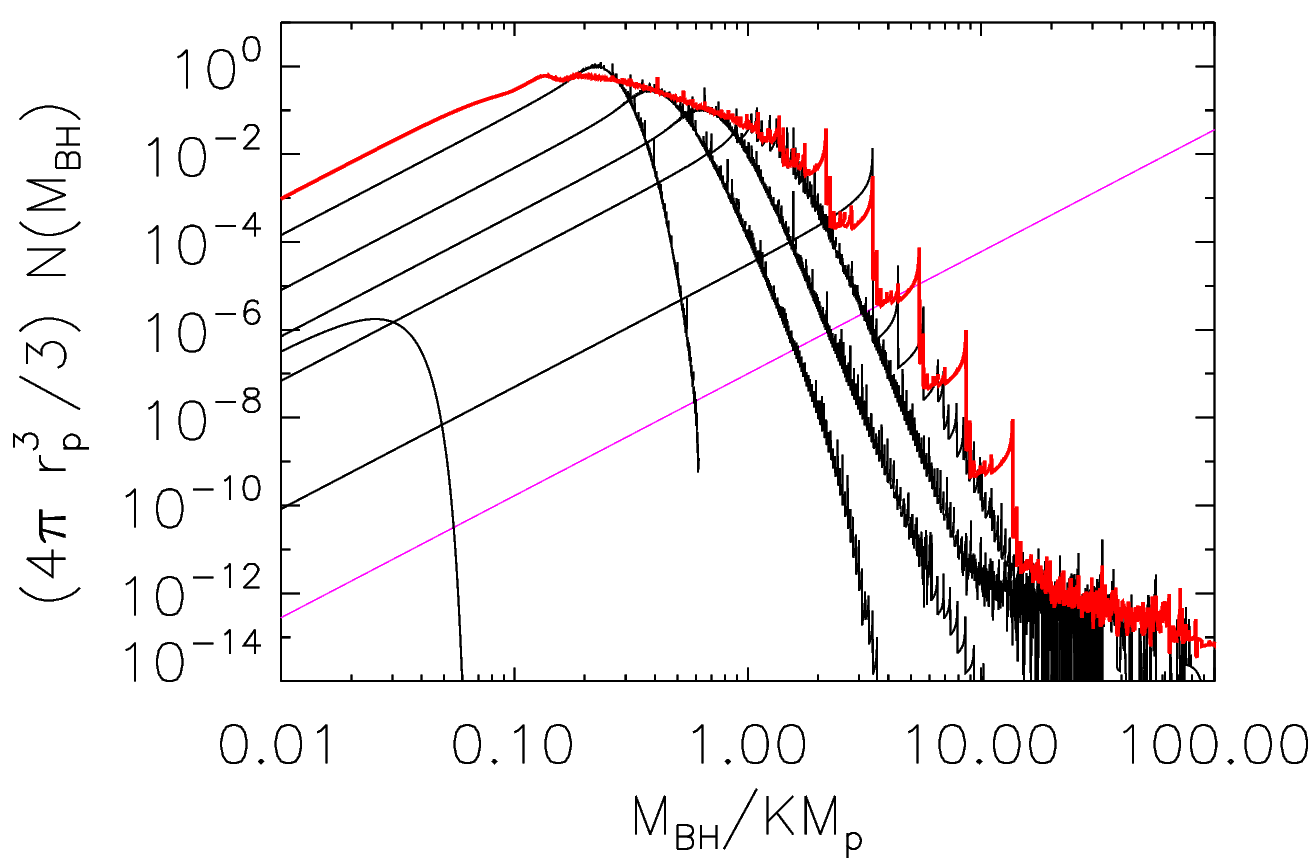} 
 \caption{Same as previous figure, but now ${\cal P}(k)$ has slope $p=2$, and we do not show any $N_{rw}$ curves.  The right hand panel has $\sigma_0 = 0.5$.  The left hand panel has the same $\sigma_0$ as for Figure~\ref{fig:fixedR0}, however the value of $A_s$ is similar to that for the right hand panel of Figure~\ref{fig:fixedR0}.  For the same $A_s$, the mass spectrum peaks at smaller $M_\bullet$ compared to when $p=0$, but the power-law slope at low masses (magenta) is the same.}
 \label{fig:fixedR2}
\end{figure}

\paragraph{Mass function:}
Figure~\ref{fig:fixedR0} shows the dimensionless scaled mass function $\tilde{N}$ when $p=0$.  The two panels show different values of $A_s$, for which $\sigma_0 = 0.33$ (left) and 0.6 (right).  In each panel, the thick red curve shows $\tilde{N}(M_\bullet)$; it is a power law at low masses (magenta curve shows a power-law of the expected slope, $1/0.36$) which is truncated exponentially at larger masses.  $\tilde{N}(M_\bullet)$ is built up over time by summing over many $\tilde{N}_r(M_\bullet)$; the solid curves show these for a few choices of $r$ (larger $r$ extend to larger $M_\bullet$). 

 At any $r$, the shape of $N_r$ is a power law at low masses with an exponentially truncation at larger masses. The amplitude of the power law part is clearly not monotonic in $r$:  neither very small nor very large $r$ contribute to the final $N(M_\bullet)$.  Since $r$ is related to the time of PBH formation, this shows explicitly that PBHs form over a range of times and, at any given time, they form with a wide spectrum of masses.  Moreover, while there is a preferred formation time for $N_r$ as $r$ changes (e.g. at the $r$ for which the amplitude of the power law is greatest) the peak mass at this time is not the same as the peak mass of the red curve ($N(M_\bullet)$):  one cannot assume a delta-function in PBH formation times, or, equivalently, a dominance of the mean profile for the compaction function.  

Each $N_r(M_\bullet)$ distribution (i.e. the mass function of  black holes formed at a given time) is built from summing over different $N_{rw}(M_\bullet)$ distributions (i.e. the mass function of  black holes formed at a given time and given compaction function curvature).  Thin curves of different styles in Figure~\ref{fig:fixedR0}, show $N_{rw}(M_\bullet)$ for a few choices of $r$ and $w$.  The cut-off properties of $N_{rw}(M_\bullet)$ can be understood as arising from the interplay between the pole in $dg_\bullet/d\ln M_\bullet$ (Eq.~\ref{eq:nmScaling}) and the exponential suppression coming from $p(g_\bullet)$; both act to modify what would otherwise be a power-law of slope $1/0.36 = 2.78$.  In particular, here we see explicitly that the exponential cutoff arises from the fact that, given $r$ and $w$, there is a maximum possible mass which is set by requiring $g_\bullet \le 4/3$.  (Note that this is slightly more stringent than just saying the mass cannot exceed that within the horizon.)  For a given $A_s$, this maximum is larger if $r$ is larger, and it is smaller if $w$ is larger (because $C_c(w)$ increases as $w$ increases).  Finally, just as one cannot assume a delta function in $r$, one cannot assume a delta-function in $w$ either.  This again shows that the use of a mean profile would give an incorrect result.

To illustrate how the shape, rather than amplitude of power spectrum affects the predictions, Figure~\ref{fig:fixedR2} shows a similar analysis of the case in which $p=2$.  The mass spectrum clearly peaks at lower $M_\bullet/M_p$ than before, and the range of $r$ that contributes signficantly is also smaller than before, but the qualitative trends remain:  a power law of slope $1/0.36$ at lower masses is truncated at larger masses, at first gradually, and then exponentially.   


\begin{figure}
 \includegraphics[width=0.475\linewidth]{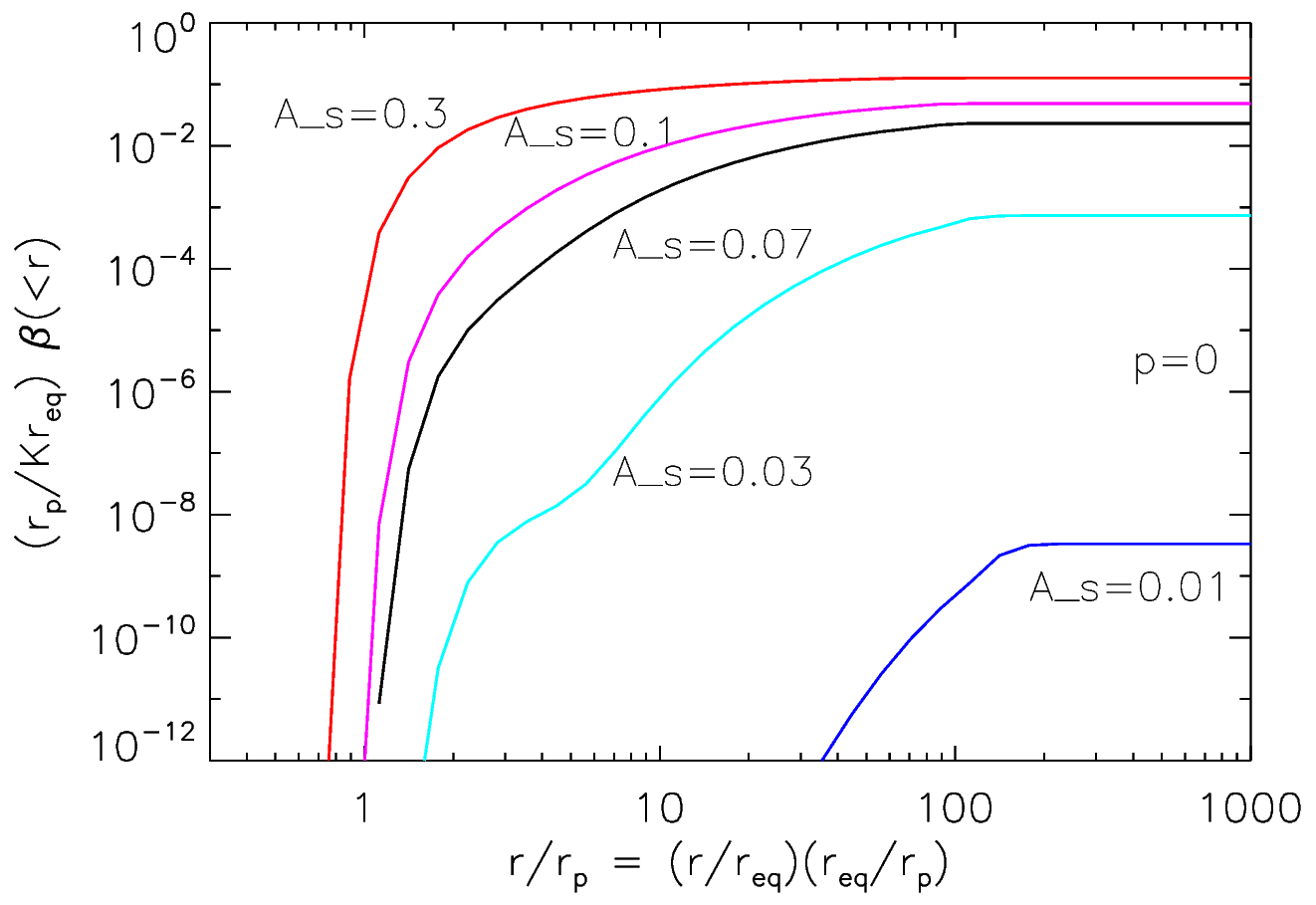} 
 \includegraphics[width=0.475\linewidth]{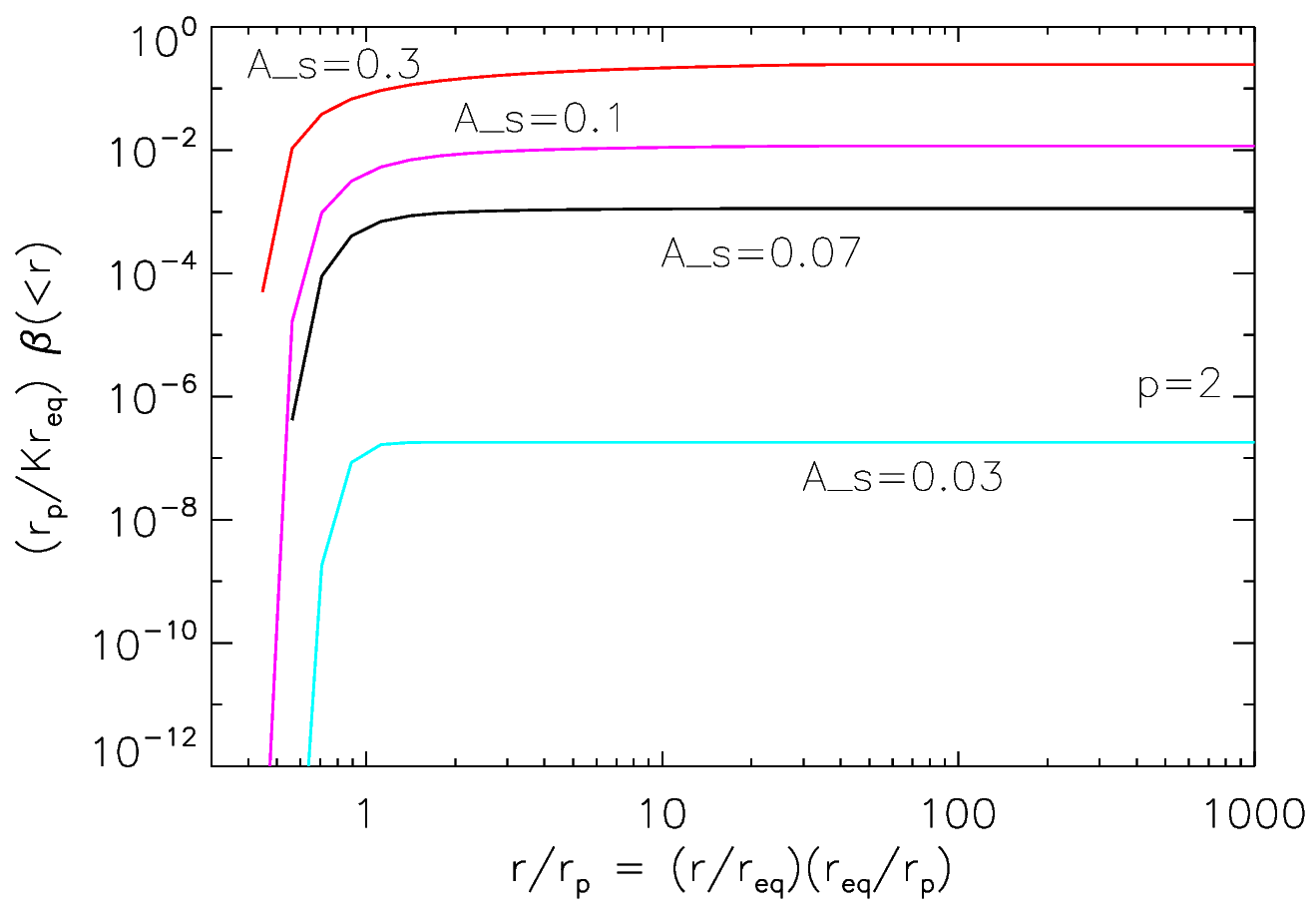} 
 \caption{Dependence of scaled mass fraction at equality (equation~\ref{tildeBeta}) that is in PBHs which formed before $r$; panels show results for ${\cal P}(k)$ having $p=0$ (left) and $2$ (right), and different curves in each panel are for different amplitudes $A_s$ (as labeled).  In both cases, this quantity is smaller if $A_s$ is smaller.  For $p=2$, most PBHs form when $r$ is slightly smaller than $r_p$; for $n=0$, the range of $r/r_p$ is broader, especially for small amplitudes.}  
 \label{fig:beta}
\end{figure}

\vspace{0.2 cm}
\paragraph{Power spectrum dependence of the abundance:}
Figure~\ref{fig:beta} shows how the predicted abundance (Eq.~\ref{eq:betaBH}) for given maximal scale $r_{max}$ depends on ${\cal P}(k)$.  The two panels are for two different ${\cal P}(k)$, and the different curves in each panel are for different amplitudes $A_s$.  Because, the horizontal-axis is a proxy for formation time, the flattening of the curves at $r\gg r_p$, is another indication that PBHs do not form at very late times.  The steepness of these curves -- how rapidly they rise to their asymptotic value -- is a measure of how narrow the distribution of PBH formation times is.  Evidently, in the panel on the right, this distribution is rather narrow, whereas in the panel on the left, it is broader.  In both panels, a smaller amplitude of ${\cal P}(k)$ results in a broader range of formation times.

Notice that, as expected, the abundances related to very small power spectrum amplitudes are exponentially suppressed. This is basically due to the fact that all variances decrease with $A_s$. Thus, for a very small power spectrum, in order to have $\beta_\bullet\sim 1$, one would need $r_p$ to be much smaller than $r_{eq}$. On the contrary, a relatively large amplitude will be related to a power spectrum peak position closer to the scale of matter-radiation equality.   

\begin{figure}
 \includegraphics[width=0.6\linewidth]{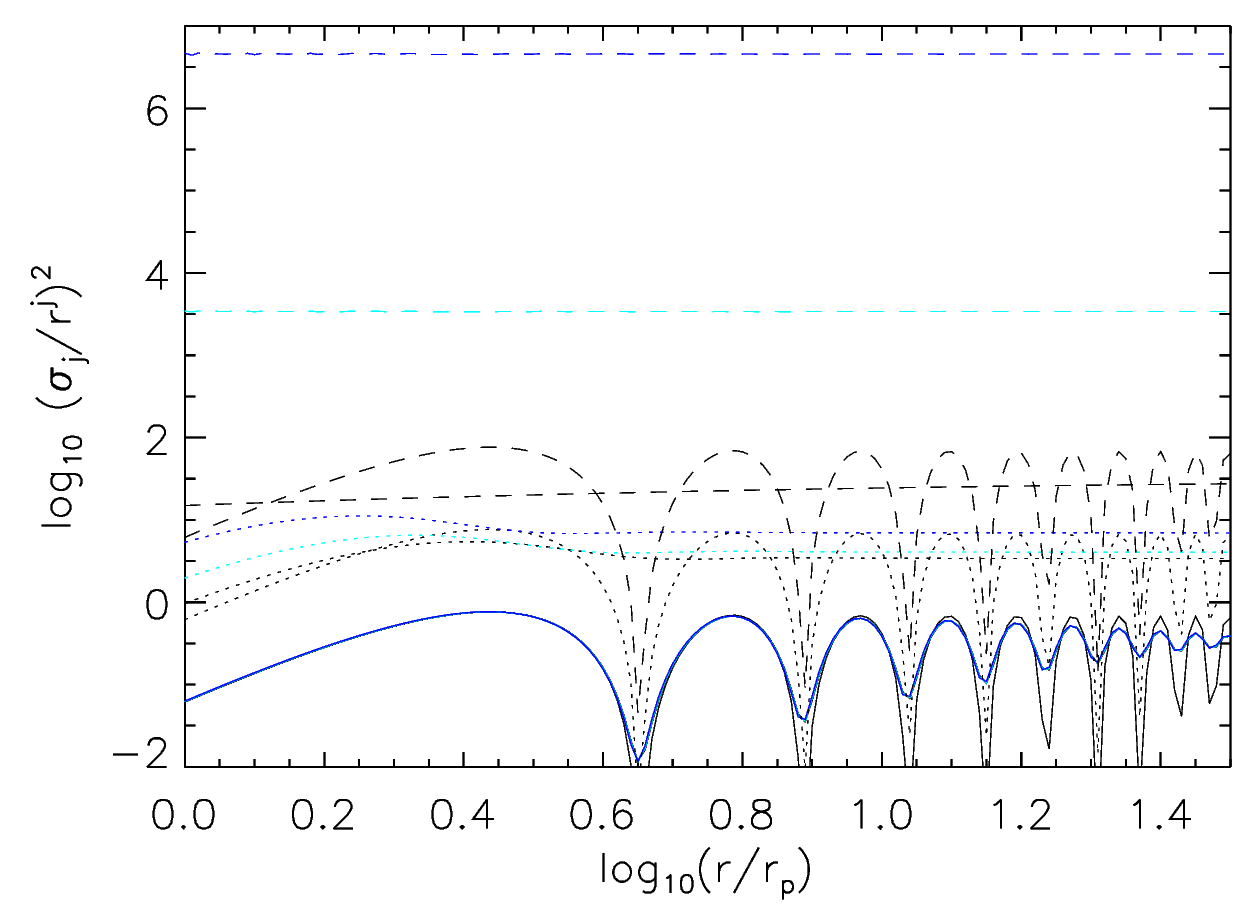} 
 \caption{Value of $(\sigma_j/r^j)^2$ in the Lognromal model for $\sigma_{\rm LN}=0.03$ (solid), $0.3$ (dotted) and 3 (dashed), for $j=0$ (black), 1 (cyan) and 2 (blue) with $A_s=1$.  Curves which drop to zero show the delta-function limit ($\sigma_{\rm LN}\to 0)$ of $\sigma_0$.  Larger $\sigma_{\rm LN}$ are just smeared versions of this limit and, at large $r/r_p$, $\sigma_j/r_j\to $ constant. }
 \label{fig:sjLN}
\end{figure}

Finally, another power spectrum shape that has appeared in the literature is the lognormal, which sets 
\be
 {\cal P}_{\rm LN}(k)=A_s\,{\rm e}^{-[\ln(k/k_p)]^2/2\sigma_{\rm LN}^2}.
\ee
Here, $\sigma_{\rm LN}$ determines the width of the feature, which is centered on $k_p$.  For small $\sigma_{\rm LN}$, the $\sigma_j$ are not monotonic functions of $r$ (see Fig.~\ref{fig:sjLN}); this can be understood by considering the delta-function limit discussed in the sharp spectrum section, and noting that $W(kr)$ oscillates (strongly!).  At larger $r$, the $\sigma_j$ in this model asymptote to the $\sigma_j\propto r^j$ scaling noted earlier.  In this respect, the model does not provide any new insights compared to the one we have already studied in this section, so we do not consider it further.
  
\section{Conclusions}
We reviewed a framework for non-linearly estimating the abundances of PBHs that were formed when radiation dominated the energy density of the Universe.  Our method explicitly differs from others in the literature: it accounts for the fact that PBHs form at positions and scales where all possible statistical realizations of compaction functions (equation~\ref{eq:C}) are maximized. Those positions and scales are intimately related to the formation time of black holes. 

Apart from predicting the abundance of PBH for any given inflationary power spectrum, we have also studied their mass distribution, the mass function. We showed that the latter is generically a power law at low masses (Figures~\ref{fig:fixedR0} and~\ref{fig:fixedR2}). The slope of this power law depends on the critical scaling law for PBH formation (equation~\ref{eq:nmScaling}) and is independent of the shape or amplitude of the underlying power spectrum of fluctuations. At larger masses there is a cutoff which depends on power spectrum shape and amplitude (Figures~\ref{fig:fixedR0} and~\ref{fig:fixedR2}).  

Our analysis shows that smaller amplitudes generically result in PBH formation that extends over a larger range in $r$ (compare top and bottom curves in left panel of Figure~\ref{fig:beta}) -- i.e., over a longer range of times.  In this respect, models which arbitrarily assume a single epoch of PBH formation related to a peak scale of the power spectrum ($r_p$) should be treated with skepticism. On the other hand, PBHs considerably lighter than $M_{\rm eq}$ (the mass of the cosmological horizon at matter-radiation equality), require $r_p\ll M_{\rm eq}^{-1}$ and a small $A_s$.  Exactly how small depends on the shape of ${\cal P}(k)$, and is the subject of ongoing work. The fact that, even with a small power spectrum one might have the whole of dark matter in PBHs is related to the fact that the mass spectrum is broader when the amplitude of the power spectrum is smaller, as can be seen Figures~\ref{fig:fixedR0} and~\ref{fig:fixedR2}. As a byproduct, this cautions against analyses which assume that PBHs all have the same mass. 

Finally we would like to remark that all our results depend critically on {\em not} including a transfer function when integrating over a power spectrum to compute statistics.  We provided a detailed discussion of why, in our setup -- in particular, because of how the compaction function is defined -- a transfer function should not be used.

\acknowledgments{}
CG is funded by the Proyecto Ministerial PID2019-105614GB-C22 and the 2021 SGR 00872 project of the Generalitat de Catalunya.  He thanks: Misao Sasaki and Shi Pi for discussions and comments on the first version of this paper, the participants of the molecule workshop ``Revisiting cosmological non-linearities in the era of precision surveys'' YITP-T-23-03, where this work was presented, the Yukawa Institute for Theoretical Physics (Kyoto, Japan) and finally the Institute of Basic Science (Daejeon, South Korea) for hospitality during the writing of this paper.
RKS thanks the ICTP for its hospitality during the summer of 2023.  Both CG and RKS thank the Institute of Cosmos Sciences for supporting the organization of the first school on Primordial Black Holes where this work was presented.


\end{document}